\documentclass[sigconf]{acmart}

\usepackage{booktabs} 
\usepackage{bbm}
\usepackage{algorithm}
\usepackage{algpseudocode}
\setcopyright{rightsretained}

\acmConference[recsysXfashion'19]
{Workshop on Recommender Systems in Fashion, 13th ACM Conference on Recommender Systems}
{September 20, 2019}
{Copenhagen, Denmark}
\acmYear{2019}
\copyrightyear{2019}
\acmArticle{}
\acmPrice{}
\acmDOI{}
\acmISBN{}

\begin{document}

\title[]{ Two-Stage Session-based Recommendations with Candidate Rank Embeddings }






\author{Jos\'e Antonio S\'anchez Rodr\'iguez}
\affiliation{%
\institution{Zalando SE}
\streetaddress{Muehlenstrasse 25}
\city{Berlin}
\state{Berlin}
\postcode{10243}
}
\email{jose.antonio.sanchez.rodriguez@zalando.de}

\author{Jui-Chieh Wu}
\affiliation{%
\institution{Zalando SE}
\streetaddress{Muehlenstrasse 25}
\city{Berlin}
\state{Berlin}
\postcode{10243}
}
\email{jui-chieh.wu@zalando.de}

\author{Mustafa Khandwawala}
\affiliation{%
\institution{Zalando SE}
\streetaddress{Muehlenstrasse 25}
\city{Berlin}
\state{Berlin}
\postcode{10243}
}
\email{mustafa.khandwawala@zalando.de}

\renewcommand{\shortauthors}{Sanchez et al.}

\begin{abstract}

    Recent advances in Session-based recommender systems have gained
    attention due to their potential of providing real-time personalized
    recommendations with high recall, especially when compared to
    traditional methods like matrix factorization and item-based collaborative filtering.
    Nowadays, two of the most recent methods are
    Short-Term Attention/Memory Priority Model for Session-based Recommendation (STAMP) and
    Neural Attentive Session-based Recommendation (NARM).
    However,  when these two methods were applied in the similar-item recommendation dataset of
    Zalando (Fashion-Similar),
    they did not work out-of-the-box
    compared to a simple Collaborative-Filtering approach.

    Aiming for improving the similar-item recommendation,
    we propose to concentrate efforts on enhancing the rank of the few most
    relevant items from the original recommendations, by employing
    the information of the session of the user encoded by an attention
    network.
    The efficacy of this strategy was confirmed when using a
    novel \textbf{Candidate Rank Embedding} that encodes the global ranking
    information of each candidate in the re-ranking process.
    Experimental results in Fashion-Similar show
    significant improvements over the baseline on Recall and MRR at 20,
    as well as improvements in Click Through Rate based on an online test.
    Additionally, it is important to point out from the evaluation that was performed
    the potential of this method on the next click prediction problem
    because when applied to STAMP and NARM,
    it improves the Recall and MRR at 20 on two publicly available real-world datasets.

\end{abstract}

%
%



\maketitle


\section{Introduction}

Recommendation systems have become a major tool for users to
explore websites with an extensive assortment. These websites offer
different recommendation products in different locations that try to
meet the specific needs of the user.

\begin{figure}
    \includegraphics[height=160pt]{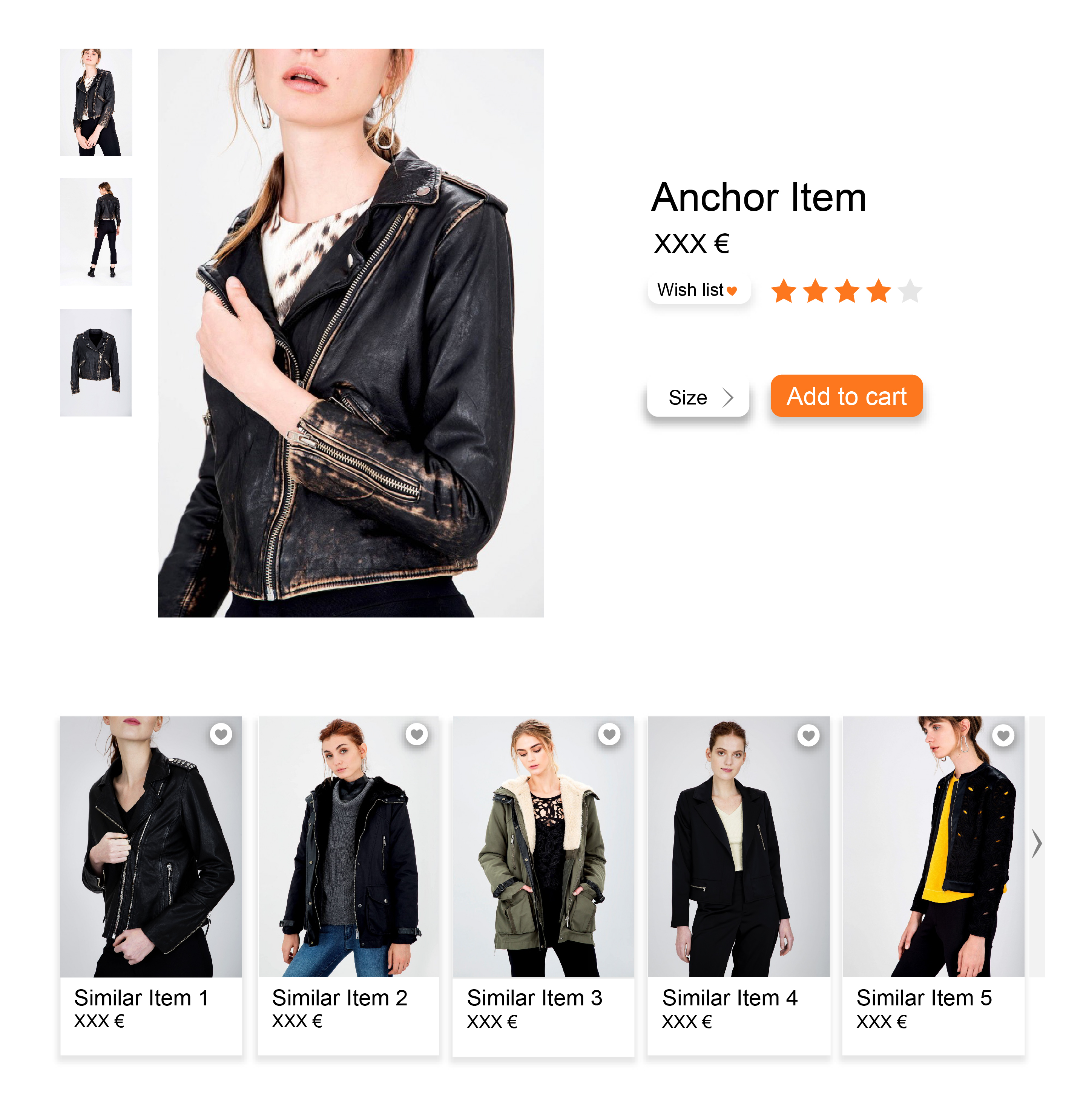}
    \caption{The product page offers similar items to the anchor.}
    \label{flg:pdp}
\end{figure}

For Zalando, one of the most important products is the similar-item
recommendation.
It is shown to all the customers that reach the product page, which offers them different alternatives related to the
product that the user is browsing. In Figure \ref{flg:pdp}, this page and the fashion-similar item
recommendation is illustrated.
In order to improve this recommendation product, a dataset that contains users' history view events and
their clicks on the recommended items was collected in early 2018 for further research.

Recent studies on Session-based Recommendations have
shown significant improvements compared to collaborative filtering approaches in several
datasets \cite{HidasiKBT15,Hidasi:2018:RNN:3269206.3271761,Tan:2016:IRN:2988450.2988452}.
Following the state-of-the-art approaches, two of the best-performing Session-based recommenders,
STAMP \cite{Liu:2018:SSA:3219819.3219950} and NARM \cite{Li:2017:NAS:3132847.3132926}
were selected to predict the next clicked item in
the similar item recommendation and compare the results with a basic collaborative filtering algorithm.
However, we were unable to obtain any improvements from the baseline in terms of recall@20.

This counter-intuitive phenomenon was also observed and reported by other studies in the literature
\cite{deSouzaPereiraMoreira:2018,DBLP:journals/corr/abs-1803-09587}.
The reasons behind the scene can be multi-faceted and complicated.
For example, one possible cause could be the bias from the feedback loop existing in the production system.
Alternatively, it could appear because of the intrinsic user behavior presented in the dataset.
Figuring out the exact reasons and applying counter-measures to fight against this phenomenon belongs to another
research area and is therefore not the focus of this paper.

On the other hand, the objective of this work is to improve the similar item recommendation in an online test.
Prior an online test, the offline evaluation metrics must be improved in order to reduce the risk of impacting
users negatively and to prioritize different online experiments.
Our hypothesis is that with the proper use of personalized information, the resulting model should be able to
outperform the baseline algorithms in both offline and online evaluations.

In order to effectively use the personalized information of user sessions, we have to overcome the performance issue
of directly applying session-based models to the Zalando Fashion-Similar dataset.
Earlier results indicate that with a single session-based model it seems to struggle in
capturing global information between items in the dataset. 
To address this issue, we propose to use Candidate Rank Embeddings (CRE) together with a session-based recommender to
collect such information from the pre-trained/pre-calculated model.

Following the approach of Covignton et. al. \cite{Covington:2016:DNN:2959100.2959190}, in the first stage a
\textbf{Candidate Generator} was employed to take a given user's click history $X_u$ as input to generate a
sorted list of candidates $C$ of size $k$.
Specifically, Collaborative filtering was used as the Candidate Generator.
In the second stage, $C$ and $X_u$ are fed to the \textbf{Re-ranker} to produce fine-tuned recommendations.
Borrowing the user session encoder of STAMP $E_{STAMP}$, $X_u$ is encoded and used in the Re-ranker together with an
innovative Candidate Rank Embeddings (CRE).
CREs learns the personalized candidate rank preference along with their item preference in the process of training
the model to optimize the prediction of the next item of interest.
By using CREs, the Re-ranker was enabled to incorporate implicit information from the Candidate Generator as prior
knowledge that helps to guide and calibrate the training of the Re-ranker with the objective of improving the
order of the original recommendation.

The offline experiments showed that our CRE-enhanced model can indeed outperform the collaborative filtering
baseline.
The good performance on the Fashion-Similar dataset suggests the CRE trick may also be applicable to other
recommendation tasks such as general next click prediction.
This hypothesis was confirmed by combining the CRE-enhanced model with two baselines, STAMP and NARM, and evaluate on
the YooChoose 1/4 and Diginetica next click prediction datasets.

The main contribution of this study lies in the following two aspects:
\begin{itemize}
    \item Candidate Rank Embeddings were used together with a session-based recommender to enhance the
    performance of I2I-CF in a two-stage approach. The model outperforms the baselines on a Fashion-Similar dataset in
    terms of Recall and MRR at 20. Also, the improvement was confirmed with an online test where it was observed
    significant improvements in Click Through Rate.

    \item Experiments and analysis were done to compare
    the baselines I2I-CF, STAMP and NARM, and the proposed method
    using those baselines as Candidate Generators on the task of predicting the next click.
    The results show that the model with CREs improves both Recall@20 and MRR@20 of these baselines on two publicly
    available datasets.
\end{itemize}

%

\section{Related Work}

\subsection{Session-based Recommender Systems}
The concept of using user interaction history as a sequence has been introduced to recommender systems with the
success of the Recurrent Neural Network (RNN) family in the sequence modeling field.
Before the thrive of these models, the user interactions were mainly used in simpler methods, such as, item-item
collaborative filtering \cite{Aiolli2013APS} or matrix factorization \cite{Koren:2009:MFT:1608565.1608614, Kumar:2014}.
One of the early attempts of using the click sequences of users as input data and considering
recommendation as a next target prediction was proposed in \cite{DBLP:journals/corr/DevooghtB16}.
Authors of DREAM \cite{Yu:2016:DRM:2911451.2914683} used pooling to summarize the current basket and later, another
vanilla RNN was used to recommend the next basket.
With the advantage of Gated Recurrent Unit (GRU) over vanilla RNN been proven for longer sequences,
GRU-based recommender systems
have been proposed with different loss functions, such as Bayesian Personalized Ranking (BPR) \cite{Rendle:2009:BPR},
TOP-1 \cite{HidasiKBT15} and their enhancements \cite{Hidasi:2018:RNN:3269206.3271761}.
These models have shown better results than matrix factorization based methods over public sequence prediction
datasets.

Various authors have proposed different improvements to the GRU family approach.
In \cite{Tan:2016:IRN:2988450.2988452} two techniques were proposed to address the data sparsity issue and combat the
behavior distribution shift problem.
In addition, the NARM architecture \cite{Li:2017:NAS:3132847.3132926} further increased the prediction power of
GRU-based session recommenders by adding an attention layer on top of the output.
Despite of the high recall and Mean Reciprocal Rank (MRR), the training process of NARM takes longer time than pure
GRU approaches.

Considering the time consumption problem of the RNN families,
authors of Short-Term Attention model \cite{Liu:2018:SSA:3219819.3219950} managed to substantially reduce the
training and prediction time by replacing the RNN structure with simpler components such as the feature average layer
and feed-forward networks.
As an improvement, the STAMP model, in addition, encodes the user feature from their click session with an attention
operation.
However, despite the promising results on several public datasets, there were studies reporting that some
state-of-the-art session-based models do not outrun simple collaborative filtering methods on a certain datasets
\cite{deSouzaPereiraMoreira:2018,DBLP:journals/corr/abs-1803-09587}.
We found the same phenomenon in our dataset generated by an online system, and that motivates us to design a solution
to have it benefit from personalization without performance degradations.


\subsection{Two-Stage Approaches}



Two-stage approaches are widely adopted in different recommendation tasks in various domains.
Paul Covington et. al. proposed the use of two neural networks to achieve video recommendation in
YouTube \cite{Covington:2016:DNN:2959100.2959190}.
With a clear separation of candidate selection and ranking generation, their solution targets especially at
multimedia recommendations. This cascading approach was used to solve performance issues, mainly due to the
enormous amount of available videos on Youtube.

Other studies have employed this technique to improve accuracy.
Rubtsov et. al. \cite{Rubtsov:2018:HTR:3267471.3267488} applied a two-stage architecture to improve the quality of
music playlist continuation.
Similar to us, they used collaborative-filtering to generate candidates and a more complicated algorithm based on
gradient boosting for the final prediction.
Likewise, we found that the use of the two-stage architecture makes it easy to improve the
model performance when applying session-based recommendations to the fashion similar item recommendation.
Besides applying a two stage approach to use the session information, we modeled the user-rank preference
with Candidate Rank Embeddings.

\section{Problem Statement}



The Session-based Recommendation problem is usually modeled as predicting the next item of interest for the user,
providing their interaction history.
Given an assortment of items $I$ where all possible items come from, the short-term history $X_u$ of a user
$u$ consists of a sequence of $l$ items $x_0, x_1, \dotsb, x_{l-1} \in I$ that $u$ has interacted with so far.
Session-based Recommenders aim to shift the next item $x_l = y_{u^+}$ that the user will interact with to the top
position of the recommendation list when given $X_u$.
A collection of user interaction sequences is denoted as a dataset $\mathcal{D}$, it is composed of $N$ pairs of user
sequence and target item $\{(X_u, y_{u^+})$, for $u=1,\, 2,\, \dotsb,\, N\}$.

A session-based recommender produces a sorted list $L_Y$ from all items in a subset $Y \subseteq I$.
In most cases, $Y$ is equivalent to $I$, but it can also come from a selected set of
candidates from another recommender.
To obtain $L_Y$, a score $s_y$ for each item $y \in Y$ is calculated and all the items from $Y$
are ranked by their scores in a descending order.
The scores of items in $Y$ is denoted as $S_Y$ and the function $rank$ yields the item
list $L_Y$ with the new order according to $S_Y$:
\begin{equation}
    L_Y = rank(S_Y)
\end{equation}

In the following sections, we use $V_x \in \mathbb{R}^d$ to represent the latent vector for an item $x$.
To represent a matrix of feature vectors of items, the notation $V_Q$ is used;
where $Q$ is a list or a set of items.
The shape of $V_Q$ is $\mathbb{R}^{d \times |Q|}$.


\section{Two-Stage Recommeder with Candidate Rank Embeddings}

\begin{figure}[hbt!]
    \includegraphics[height=80pt]{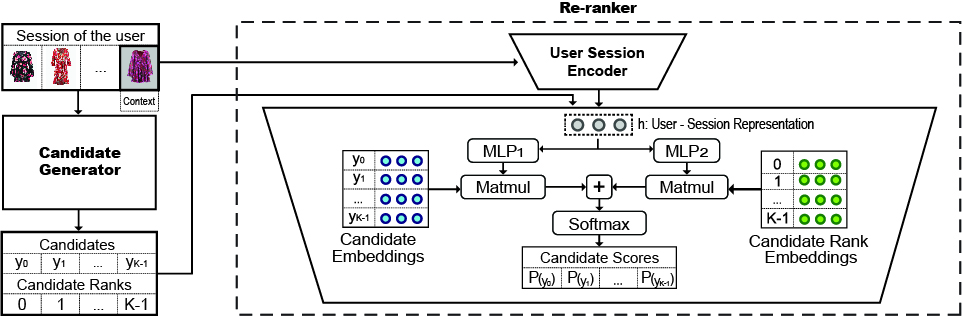}
    \caption{The model architecture: The Candidate Generator is trained first and treated as a black box.
    Then, the Re-ranker is trained to re-score the candidates provided by the Generator.
    For calculating these scores, two components are considered: the candidates and the rank preference of them.}
    \label{flg:model_arch}
\end{figure}

Using the naming convention from \cite{Covington:2016:DNN:2959100.2959190}, the first recommender of the cascade
is considered as the \textbf{Candidate Generator} $\mathbf{G}$, while the second,
known as the \textbf{Re-ranker} $\mathbf{R}$, ranks the most relevant items from the output
of $\mathbf{G}$.
Both $\mathbf{G}$ and $\mathbf{R}$ take a candidate set and a specific user history as input parameters,
and later assign a score to each candidate.
The final recommendation of the proposed method $L_Y$ is calculated as follows:
\begin{equation}
    \begin{aligned}
        & L_{Y_G} = rank(S_{Y_G}) = rank(G(Y,\, X_u,\, \theta))\\
        & C,\,\, \overline{C} = L_{Y_G}[:k],\,\, L_{Y_G}[k:] \\
        & L_{C_R} = rank(S_{C_R}) = rank(R(C,\, X_u,\, \mathit{w})) \\
        & L_Y = L_{C_R} : \overline{C}
    \end{aligned}
    \label{defineGandR}
\end{equation}
Where $C$ denotes the $k$ most relevant candidates computed by $G$ and $\overline{C}$ indicates the rest in $Y$.
$\theta$ and $\mathit{w}$ denotes the trainable parameters of $G$ and $R$ respectively.
In this study, we set $\theta$ and $\mathit{w}$ to be independent and do not share parameters.
The training process of the proposed model is also two-staged.
$G$ is trained first (in case of training needed).
The training of $R$ starts after a well-trained $G$ is obtained, and only considers the top-ranked $k$ candidates coming
from $G$, as described in Eq \ref{defineGandR}.
The parameters $\theta$ and $\mathit{w}$ are optimized using $\mathcal{D}$.
Figure \ref{flg:model_arch} illustrates the overall architecture of the model.

At inference time, given a specific user interaction history $X_u$,
both, $G$ and $R$, take it as an input and operate sequentially.
More details about the Candidate Generator and the Re-ranker are described in
Sections \ref{candidate_generator} and \ref{reranker}


\subsection{The Candidate Generator} \label{candidate_generator}
The Candidate Generator $G$ can be an arbitrary recommender that takes a user session $X_u$ as input and ranks the
set $Y_G$. For training $G$ we consider $Y_G = I$.

The selection of the algorithm for $G$ depends mostly on the characteristics of the dataset $\mathcal{D}$
and the performance of the algorithm on generating high quality candidates.

\subsection{The Re-ranker} \label{reranker}
The Re-ranker $R$ takes the same user click sequence $X_u$, as $G$ does, but concentrates on ranking a smaller set of
candidates $C$ determined by the Candidate Generator it connects with.

We employ a variant of STAMP in $R$ to include the candidate rank information in the model.
Specifically, the encoder for the user click history $E_{STAMP}$ is reused.
$E_{STAMP}$ is a simple element-wise multiplication between the history representation $h_s$ and the anchor
representation item $h_t$ described in \cite{Liu:2018:SSA:3219819.3219950}.
\begin{equation}
    \begin{aligned}
        & h_u = E_{STAMP}(X_u) = h_s \odot h_t\\
    \end{aligned}
    \label{stamp_encoder}
\end{equation}
The STAMP encoder $E_{STAMP}$ was chosen because of its architectural simplicity and promising performance in
predicting the user preference.

Before defining the calculation of the scores, Candidate Rank Embeddings must be introduced.
Given a sequence $X_u$, we obtain a sorted list of candidates $C = [y_0, y_1, \dotsb , y_{k-1} ]$ of size $k$ from $G$.
The \textbf{Candidate Rank} $r_i$ is an integer ranging from $[0, k)$ which denotes the position of the $i^{th}$ item
in $C$.
Each Candidate Rank is associated with a Candidate Rank Embedding.
Therefore, CREs are positional embeddings shared among different candidate lists produced by $G$.
The CREs for a Re-ranker that takes $k$ candidates into account can be represented as a matrix $W_{CR}$ of shape
$\mathbb{R}^{k \times d_{CRE}}$.

With the CRE defined the scores are specified in the following equation:
\begin{equation}
    \begin{aligned}
        & h_e = MLP_1(h_u) = tanh(W_{e1}^\intercal{tanh}(W_{e2}^\intercal{h_u} + b_{e2}) + b_{e1}) \\
        & h_r = MLP_2(h_u) = tanh(W_{r1}^\intercal{tanh}(W_{r2}^\intercal{h_u} + b_{r2}) + b_{r1}) \\
        & D_{R}(h_u,\, C) = softmax(V_{C}^\intercal{h_e} + W_{CR}^\intercal{h_r})
    \end{aligned}
    \label{dreranker}
\end{equation}

Where $W_{e1}$, $b_{e1}$, $W_{e2}$, $b_{e2}$, $W_{r1}$, $b_{r1}$, $W_{r2}$, $b_{r2}$ are learnable weight matrices
and bias terms of the feed-forward network layers, $V_{C}$ is the item embeddings of the candidates and
$W_{CR}$ is the candidate rank embedding matrix.
Note that $W_{CR}$ is the same for all user sequences $X_u$ because they depend only on the rank of the
candidates.
All the embeddings are initialized randomly and trained together with the model.

We train $R$ to predict the next click that is in $C$ by using the Cross Entropy loss.

The main difference between our solution and STAMP is the use of $W_{CR}$ and two non-linear projections $MLP_1$ and
$MLP_2$.
These projections are used to approximate the click probability of the candidates and the rank preference of users.
The first projection focuses on predicting the embedding of the target item, while the second one
focuses on predicting the embedding of the position of the target item in the ranked candidate list.

The intuition behind learning the rank preference is that the information from the output of $G$ can flow into the
model, and a balance between the newly-learned item preference and the old rank can be obtained by summing up two user
preferences.

Furthermore, since the ranking score comes from the dot product from the candidate rank embeddings and a projection
from the user representation, this allows the model to learn the relationship between the user and the
position of the target. For example, it gives the capability to recognize which type of
users like to click the top positions or the items which co-occur with the anchor very often when $G$ produces its
candidates using co-occurrence information.
This behavior can be difficult to learn with a model that considers only the user-item preference.

In Section \ref{analysis} an analysis of the importance of using the candidate rank information is presented,
where we compare our two-stage approach against one without CREs.

So far, we only tried using a one-to-one mapping between Candidate Ranks/positions and CREs.
In applications with a large candidate set, having multiple ranks share one CRE could be beneficial
because training signals can be shared among several unpopular positions.

\section{Experiments and Analysis}

\subsection{Baselines}
The following recommendation algorithms are used in our experiments as the baselines.
\begin{itemize}
    \item \textbf{Item-Item Collaborative Filtering (I2I-CF)}: it considers only the most
    similar items for the last-seen item $x_{l-1}$ in the user interaction/click sequence. It is
    pre-calculated using a variant of the cosine similarity function described in \cite{Aiolli2013APS}.
    \item \textbf{Attention-based GRU (NARM)} \cite{Li:2017:NAS:3132847.3132926}
    \item \textbf{Short-Term Attention Model (STAMP)} \cite{Liu:2018:SSA:3219819.3219950}
\end{itemize}

\subsection{Experiment Setup}
Each of the baselines is used as a Candidate Generator in two sets of separated experiments.
The first set of experiments evaluate the model performance on the Fashion-Similar dataset to predict the next
similar item on the carousel that the user would click.
The second set of experiments compare the performance of CREs on two next item prediction datasets, when combining
with different baseline models.
Before training the re-ranker, we first train the corresponding baseline and keep them fixed as Candidate Generators.
Not all the training sequences are used in training, the re-ranker only considers
those which their target item falls within the candidate set $C$ from $G$.
Five percent of the training examples are randomly sampled for validation.
During training, the model performance is checked every 1000 steps and the best model is selected by their performance
in terms of Recall@5.
Adam \cite{DBLP:journals/corr/KingmaB14} is used to train $R$ for $5$ epochs with a learning rate
of \textbf{0.001} and batch size of \textbf{$512$}.
We set the number of candidates $k$ being returned by $G$ to be \textbf{100}.
The item embeddings and model weights of $E_{STAMP}$ are initialized w.r.t. the best settings reported
in \cite{Liu:2018:SSA:3219819.3219950}.
The weights of $W_{CR}$, $W_{e1}$, $W_{e2}$, $W_{r1}$, $W_{r2}$, are initialized with Xavier \cite{pmlr-v9-glorot10a}.
To label the results, we use RRCRE-X as an abbreviation of re-ranking the output from an approach X with CREs.

\subsection{Predicting the Fashion Similar Item}

In this task, our goal is to predict the clicked products in the similar-item-recommendation carousel, given the latest
$l$ actions of a user.

Only the \textbf{Fashion-Similar} dataset was used because the other datasets are mainly used for general next click
prediction tasks.
The dataset was collected using the customers actions from several major european markets for multiple days.
Every record in this dataset consists of a click on the similar-item recommendation, which is the target,
along with the latest 12 items the user browsed before interacting with the similar-item recommendation.

The dataset was split into training and test sets by time, leaving 9 days for the training set and the last day for
the test test.
With this arrangement, the offline evaluation measures how well the model performs on the next day assuming we retrain
our algorithms daily.

The training set contains 8353562 examples, for 1435605 users which interacted with 650228 items in total.
The test set contains 624559 examples, for 233051 users which interacted with 259784 items in total.

\begin{table}
    \caption{Performance of the baselines and the proposed method on predicting the similar item.}
    \label{table:result_similar}
    \begin{tabular}{cccl}
        \toprule
        &\multicolumn{2}{c}{Fashion-Similar} \\
        \midrule
        & Recall@20 & MRR@20\\
        \midrule
        I2I-CF & \textbf{0.8106} & \textbf{0.2611} \\
        STAMP & 0.7244 & 0.2443  \\
        NARM & 0.6989 & 0.2510  \\
        \midrule
        RRCRE-CF & \textbf{0.8381} & \textbf{0.2981} \\
        \bottomrule
    \end{tabular}
\end{table}

\subsection{Predicting the next click}
For this task, the objective is to predict the next user interaction given the past click history.
We used the following datasets:
\begin{itemize}
    \item \textbf{YooChoose 1/4}: the dataset was preprocessed exactly as described in \cite{Tan:2016:IRN:2988450.2988452}.
    \item \textbf{Diginetica}: we preprocessed the dataset with the approach described in \cite{Wu:2019vb}. Note that Wu
    et. al. introduced additional preprocessing to the dataset compared to \cite{Li:2017:NAS:3132847.3132926}.
\end{itemize}


\begin{table}
    \caption{Performance of the baselines and the proposed method on the task of predicting the next click.}
    \label{table:result_next_click}
    \begin{tabular}{cccccl}
        \toprule
        & \multicolumn{2}{c}{YooChoose 1/4} & \multicolumn{2}{c}{Diginetica} \\
        \midrule
        & Recall@20 & MRR@20 & Recall@20 & MRR@20\\
        \midrule

        I2I-CF & 0.5259 & 0.2001 & 0.3760 & 0.1211\\
        STAMP & \textbf{0.6983} & 0.2915 & 0.4834 & 0.1588 \\
        NARM & 0.6973 & \textbf{0.2921} & \textbf{0.5015} & \textbf{0.1599} \\

        \midrule
        RRCRE-I2I-CF & 0.5586 & 0.2446 & 0.3773 & 0.1220\\
        RRCRE-STAMP & \textbf{0.7086} & \textbf{0.3133} & 0.5046 & \textbf{0.1677} \\
        RRCRE-NARM & 0.7029 & 0.3082 & \textbf{0.5116} & 0.1675 \\
        \bottomrule
    \end{tabular}
\end{table}





\subsection{Offline Results and Analysis}
\label{analysis}
%
%

From Table \ref{table:result_similar} we can observe that STAMP and NARM don't show superior results in the
task of predicting the similar item in the offline evaluation.
The result on Fashion-Similar seems to be counter-intuitive, and further investigations are required as future work.

However, with RRCRE-I2I-CF we are able to improve I2I-CF in the similar item prediction task in terms of Recall@20 and
MRR@20.
It is because the model is capable of utilizing the hidden information in the ranking of the baseline together with
the session information captured by the attention network from $E_{STAMP}$.
An online test was performed to confirm the offline results, described in Section \ref{online}.

As shown in Table \ref{table:result_next_click}, STAMP and NARM perform significantly better than I2I-CF in the next
click prediction task.

We also applied our method to use STAMP and NARM as Candidate Generator in the next click prediction task, the
evaluation result shows that it is able to slightly improve the Recall@20 and MRR@20 of STAMP and NARM on both
YooChoose 1/4 and Diginetica.

\begin{figure}
    \includegraphics[width=0.5\textwidth, clip]{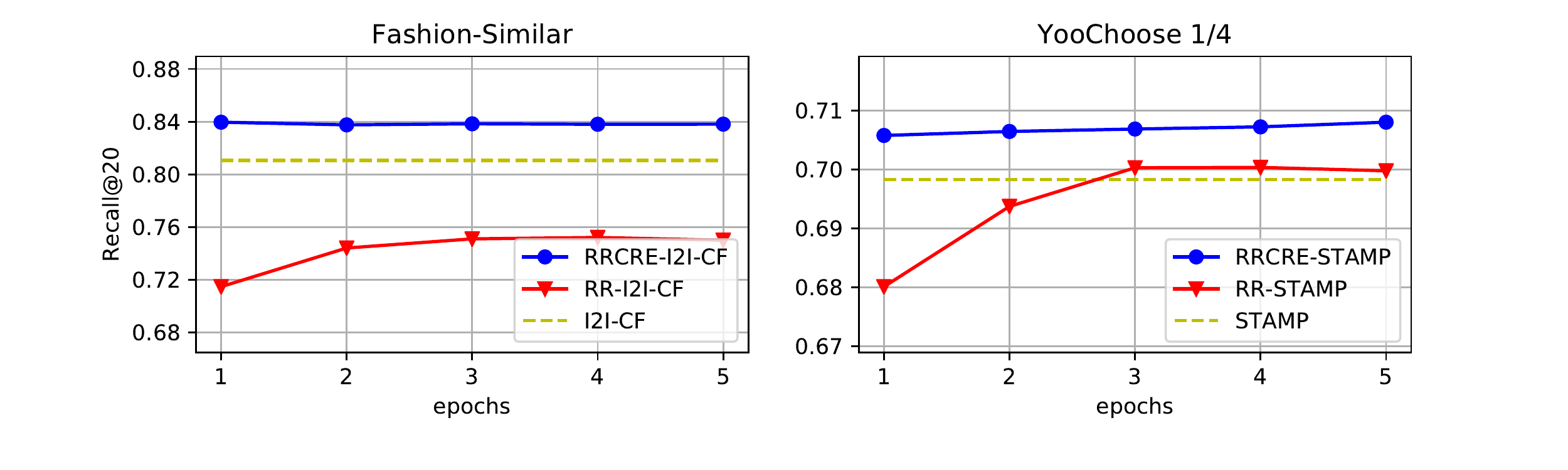}
    \caption{
    Our approach with and without Candidate Rank information.
    }
    \label{fig:wp_comparison}
\end{figure}

To understand the improvement obtained by applying the Candidate Rank Embeddings, we compared the model performance
between the proposed method with and without CRE, i.e. the final scores of candidates
become $D_{R}(h_u,\, C) = softmax(V_{C}^\intercal{h_e})$.
We use RR-X as an abbreviation of re-ranking the output from an approach X without CREs.
The result is illustrated in Figure \ref{fig:wp_comparison}.
For the Fashion-similar dataset, we can observe that simply re-ranking the most relevant candidates from I2I-CF
with $E_{STAMP}$ doesn't lead to superior results.
On the other hand, when training with CREs we obtain a better result listed in Table \ref{table:result_similar}.
We also compare RR-STAMP and RRCRE-STAMP in one of the next click prediction datasets.
It turns out that RRCRE-STAMP outperforms the baseline from epoch 1, while RR-STAMP requires more iterations.
It is because RR-STAMP has to learn the next clicked target from randomly-initialized model parameters without the rank
information from STAMP being presented.

%
%
\begin{figure}
    \includegraphics[width=0.5\textwidth, clip]{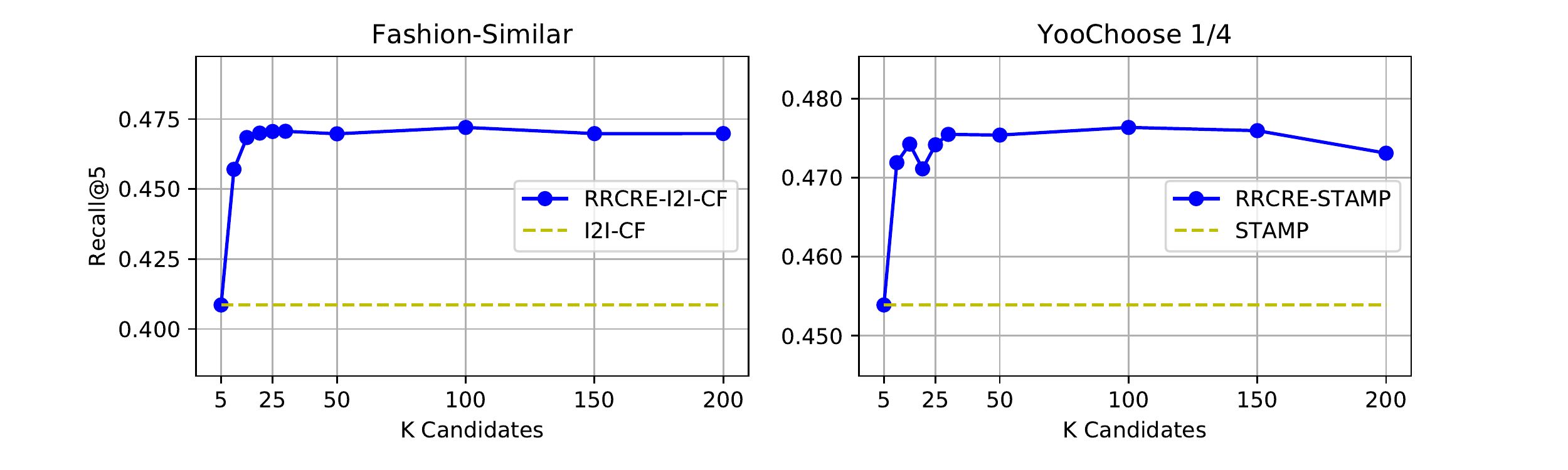}
    \caption{Recall@5 of the bests approaches on each dataset given different number of candidates to re-rank.}
    \label{fig:k_candidates}
\end{figure}

Additionally, we illustrate the behaviour of the proposed method with respect to the number of candidates to re-rank.
In Fig \ref{fig:k_candidates} we can observe improvements in recall@20 even with a small $k$.
In addition, we found the recall plateaus or decreases when $k$ exceeds a certain threshold.
One possible reason could be that the candidates associated with the low ranks rarely appear as targets in the dataset.
As a result, the CREs for these ranks could not have been well-trained and could have captured misleading information.
However, since using a relatively small $k$ simplify the effort of training and serving in production, more
investigations were not done.




\subsection{Online test in Zalando}
\label{online}

From the previous section, it was observed an improvement in the offline metrics with respect to I2I-CF.
As a consequence, an online test hosted by Zalando was performed to compare I2I-CF and RRCRE-I2I-CF.

RRCRE-I2I-CF was served using cpu machines running Tensorflow-Serving.
For this algorithm, the recommendations are calculated in real-time.
On the contrary, I2I-CF was served by using a static table stored in memory.

I2I-CF was giving static non-personalized similar recommendations and RRCRE-I2I-CF was adapting the similar item list
depending on the user's previous $l$ actions.

The models were updated every day to be adapted to the latest user behaviour and we compared their
performance in major european markets for several days.

To ensure that the recommendations satisfy the similarity constraint of the product,
some filters based on category trees were applied to the output of both methods.

The results showed relative improvements in engagement based on a significant $\textbf{+2.84\%}$ ($p-value \leq 0.05$)
increase in Click Through Rate.
It proves that there is a positive effect of using the session of users to generate a personalize
ranking of similar items and supports the offline experiment results carried-out.

\section{Conclusion and Future Work}

In this study, the possibility of improving
the fashion similar item recommendation was explored with a two-staged
re-ranking approach that is able to benefit from the Candidate Rank information,
the session of the user and a small set of candidates.

With this approach, the Recall@20 and MRR@20 of I2I-CF was improved on the Fashion-Similar dataset, and the
success in the offline evaluation was confirmed by an online test.

The proposed approach was also confirmed to be useful to improve the performance of two advanced session-based
recommendation algorithms, STAMP and NARM on the next click prediction datasets YooChoose 1/4 and Diginetica.
Despite the success in the offline evaluation, further experiments are needed to confirm the impact of the proposed
method in the context of session-based recommendation.


\begin{acks}
    The authors are immensely grateful to
    Alan Akbik,
    Andrea Briceno,
    Humberto Corona,
    Antonino Freno,
    Francis Gonzalez,
    Romain Guigoures,
    Sebastian Heinz,
    Bowen Li,
    Max Moeller,
    Roberto Roverso,
    Rezar Shirvany,
    Julie Sanchez,
    Hao Su,
    Lina Weichbrodt and
    Nana Yamazaki
    for their support, revisions, suggestions, ideas and comments that greatly helped to improve the quality of this
    work.

\end{acks}

\bibliographystyle{ACM-Reference-Format}
\bibliography{bibliography}

\end{document}